\documentclass{IEEEtran}
\usepackage{cite}
\usepackage[T1]{fontenc}
\usepackage{amsmath, amsthm, amssymb}
\usepackage[flushleft]{threeparttable}
\usepackage{array,booktabs,makecell}
\usepackage[ansinew]{inputenc}
\usepackage{hyperref}
\usepackage{xcolor}
\hypersetup{
    colorlinks=true,
    linkcolor=black,
    filecolor=black,      
    urlcolor=black,
    citecolor=black
}
\usepackage{amsmath,amssymb,amsfonts}
\usepackage{graphicx}
\usepackage{textcomp}
\usepackage{xcolor}

\usepackage{mathtools}

\DeclarePairedDelimiter\abs{\lvert}{\rvert}%
\DeclarePairedDelimiter\norm{\lVert}{\rVert}%

\makeatletter
\let\oldabs\abs
\def\abs{\@ifstar{\oldabs}{\oldabs*}}
\let\oldnorm\norm
\def\norm{\@ifstar{\oldnorm}{\oldnorm*}}
\makeatother

\begin{document}
\title{\huge{Critical Temperature Prediction for a Superconductor:\\
A Variational Bayesian Neural Network Approach}}

\author{Thanh Dung Le, \IEEEmembership{Member, IEEE}, Rita Noumeir, \IEEEmembership{Member, IEEE}, Huu Luong Quach, Ji Hyung Kim, \\
Jung Ho Kim, and Ho Min Kim, \IEEEmembership{Member, IEEE}

\thanks{Manuscript received Sep 24, 2019; accepted Jan 29, 2020. Date of
publication Jan 29, 2020; date of current version Jan 29, 2020. The author (T. D. Le) acknowledges the financial support of the Canada First Research Excellence Fund (CFREF) program through IVADO, and the Doctoral Scholarship for International Student from Le Fonds de Recherche du Quebec Nature et technologies (FRQNT).}
\thanks{T. D. Le and R. Noumeir are with the Biomedical Information Processing Lab, Ecole de Technologie Superieure, Montreal, QC H3C 1K3, Canada (Corresponding email: thanh-dung.le@etsmtl.ca). }
\thanks{H. L. Quach, J. H. Kim, and H. M. Kim are with the Applied Superconducting Lab, Jeju National University, Jeju-si 690-756, S. Korea (Email: \{qhluong, jihkim, hmkim\}@jejunu.ac.kr).}


\thanks{J. H. Kim is with the Institute for Superconducting \& Electronic Materials, Australian Institute of Innovative Materials, University of Wollongong, Wollongong NSW 2522, Australia (Email: jhk@uow.edu.au).}
}
\markboth{IEEE TRANSACTIONS ON APPLIED SUPERCONDUCTIVITY: ACCEPTED FOR PUBLICATION}{SKM: My IEEE article}

\maketitle

\begin{abstract}
Much research in recent years has focused on using empirical machine learning approaches to extract useful insights on the structure-property relationships of superconductor material. Notably, these approaches are bringing extreme benefits when superconductivity data often come from costly and arduously experimental work. However, this assessment cannot be based solely on an open black-box machine learning, which is not fully interpretable, because it can be counter-intuitive to understand why the model may give an appropriate response to a set of input data for superconductivity characteristic analyses, e.g., critical temperature. The purpose of this study is to describe and examine an alternative approach for predicting the superconducting transition temperature $T_c$ from SuperCon database obtained by Japan's National Institute for Materials Science. We address a generative machine-learning framework called Variational Bayesian Neural Network using superconductors chemical elements and formula to predict $T_c$. In such a context, the importance of the paper in focus is twofold. First, to improve the interpretability, we adopt a variational inference to approximate the distribution in latent parameter space for the generative model. It statistically captures the mutual correlation of superconductor compounds and; then,  gives the estimation for the $T_c$. Second, a stochastic optimization algorithm, which embraces a statistical inference named Monte Carlo sampler, is utilized to optimally approximate the proposed inference model, ultimately determine and evaluate the predictive performance. As a result, in comparison with the standard evaluation metrics, the results are promising and also agree with the existing models prevalent in the field. The $R^2$ value obtained is very close to the best model (0.94), whereas a considerable improvement is seen in the RMSE value (3.83 K). Notably, the proposed model is known as the first of its kind for predicting a superconductor's $T_c$.
\end{abstract}

\begin{IEEEkeywords}
Critical transition temperature, machine learning, Bayesian neural network, variational inference, stochastic optimization algorithm, high temperature superconducting (HTS).
\end{IEEEkeywords}

\section{Introduction}
\label{sec:introduction}
\IEEEPARstart{T}{he} generality of machine learning (ML) in material science is increasingly being adopted to discover hidden trends in data and make predictions. Mainly, there are guidance and perspectives when applying ML techniques as a robust protocol to maintain both quantitatively and qualitatively predictive models \cite{wagner2016theory}; impacts of ML technologies on materials,  process  and  structures engineering are likely transformational for advancing new solutions to the long-standing data structure challenge \cite{dimiduk2018perspectives}; reliable and explainable ML models from underrepresented materials data provide both model-level and decision-level explanations \cite{kailkhura2019reliable}. Besides, there are also wide ranges of ML applications in material data science, such as illustrative examples of a taxonomy of ML capabilities in soft mater, and data-driven materials design engines \cite{ferguson2017machine}.\vspace{2.5pt}\\ 
\indent In superconductivity, machine learning-guided iterative experimentation may outperform standard high-throughput screening for discovering breakthrough materials in high temperature (high-$T_c$) superconductors, e.g., a new measure of machine learning model performance in high-$T_c$ by improving the cross-validation with a single neural network \cite{meredig2018can}, empirical analyses for critical current measurement by developing machine learning tool in classification and regression tasks for SuperCon database \cite{stanev2018machine}. These studies confirm that the framework for making machine-based decisions and actions using ML analysis has dominated the predictive models. However, there are growing concern steps for attaining autonomous prediction that require at least three concurrently operating technologies: i) making analyses by endorsing perception of information, ii) predicting the sensed field changing over time, and iii) establishing a policy for a machine to take unsupervised action.\vspace{2.5pt}\\
\indent To address the challenges, we describe an alternative ML approach called the generative neural network model. Also, Bayesian-based generative model prediction is advantageous for the two most important reasons:
i) uncertainty is intrinsically described, useful for analysis, and prediction, ii) overfitting is avoided by natural penalization of overly
complicated models. In this work, we develop the probabilistic high-$T_c$ predictive model using Variational Bayesian Neural Network (VBNN) regression provided by Drugowitsch \cite{drugowitsch2013variational} and build upon the efficient optimization algorithm for learning optimal learnable parameters in the VBNN. In particular, we exploit the Stochastic Gradient Variational Bayes (SGVB) optimization algorithm introduced in \cite{kingma2013auto}. The learning algorithm provides the probability that latent correlation of parameters in superconductors  chemical  characteristics in drawing the $T_c$ prediction, instead of a fixed linear regression with uncertainty.\vspace{2.5pt}\\
\indent The rest of this paper is organized as follows. Section \ref{sec:work} discusses strength, weaknesses, and achievements of the related work, and Section \ref{sec:model_pre} presents the mathematical underpinning of the VBNN model and preliminary backgrounds of an inference model. Section \ref{sec:temp_model} provides a conceptual $T_c$ prediction method by applying the VBNN model and prediction evaluation. In Section \ref{sec:result}, we numerically evaluate its performance. Finally, Section \ref{sec:conclusion} provides concluding remarks.
\section{Related Work}
\label{sec:work}
One of the major impediments to the industrial take-up of high-temperature superconductors is the paucity of comprehensive, reliable, and relevant performance data on commercially available wires such as wire performance database from the Robinson Research Institute \cite{wimbush2016public}, SuperCon database maintained by Japan's National Institute for Materials Science \cite{supercond}. To address this, the article first reviews the data analytic approaches on critical temperature prediction, which is followed by a review of machine learning-based prediction models. For each study, the present paper discusses the strength, weaknesses, and achievements of different approaches, so as to better motivate the technique proposed in this work.\vspace{2.5pt}\\
\indent Recently, an active collaboration of Center for Nanophysics and Advanced Material members, together with researchers from the National Institute of Standards and Technology and Duke University, has been exploring methods by developing machine learning schemes to model the $T_c$ of over 12,000 known superconductors  published by Nature \cite{stanev2018machine}. All machine learning in this work are variants of the random forest method. Random forest is used in this work because of several advantages, such as i) can learn complicated non-linear dependencies from the data, ii) quite tolerant to heterogeneity in the training data , iii)  can estimate the importance of each tree predictors, thus making the model more interpretable. The results show that a single regression model combined with a backward feature elimination process did a reasonably well; the model achieved statistical measure $R$-squared ($R^2) \approx 0.85$.\vspace{2.5pt}\\
The study \cite{hamidieh2018data} also applies random forest to predict the superconducting critical temperature based on the features extracted from the superconductor's chemical formula. However, this study enhanced the random forest by adapting XGboost (eXtreme Gradient Boosting) approach. The gradient boosted models create an ensemble of trees to predict a response. The trees are added sequentially to improve the model by accounting for the points which are difficult to predict. Then, gradient boosted models can handle the complex interactions and correlations among the features. Consequently, the regression model serves as a benchmark with 17.6 K and 0.74 for the out-of-sample root mean square error (RMSE) and $R^2$.\vspace{2.5pt}\\
\indent Early, the work \cite{owolabi2016application} directly relates the lattice parameters to the $T_c$ through computational intelligence technique using support vector machine (SVM) regressor. The success of the model paves a significant way for quick and accurate estimation of $T_c$ of doped YBCO superconductors, and eventually eases the usual high demanding experimental procedures that involve the use of expensive cryostat. Technically, SVM employs a mapping function called kernel function to map non-linear regression to high dimensional feature space where linear regression is conducted. Then, they will find the optimal value by tuning the training of all parameters from the SVM problem. The results of the modeling and simulations indicate that the developed approach was capable of estimating the $T_c$ with a high degree of accuracy as can be deduced from high coefficients of correlation of 96.65\% and 95.75\% during the training and testing periods of the model, respectively.\vspace{2.5pt}\\
\indent The recent exciting study \cite{konno2018deep} holds a promising technique to predict new high-temperature superconductors. The author introduced a novel approach using deep learning and succeeded in making a new list of candidate materials of superconductors. They proposed a new approach ``material as images in periodic table.'' Technically, elements and materials are presented as images in the periodic table because number of electrons and their orbit determine the properties of the elements. Then, a deep convolutional neural network (CNN) is adapted for images from the periodic table with four channels corresponding to $s$-block, $p$-block, $d$-block, and $f$-block. The results confirmed the accuracy 94\%, precision 74\%, the recall 62\% and $f_1$ score is 67\% with the SuperCon data. The achieved $R^2$ of the critical temperature in the test data is 0.93.\vspace{2.5pt}\\
\indent To confirm the promise of the study \cite{konno2018deep}, another study recently published in Nature \cite{zeng2019atom} develops an atom table CNN atom table  CNN learning the experimental properties directly  from the features constructed by itself. In particular, CNN only requires the component information. Through data-enhanced technology, their model not only accurately predicts superconducting transition temperatures, but also distinguishes superconductors and non-superconductors. Utilizing the trained model, they have screened 20 compounds that are potential superconductors with high superconducting transition temperature from the existing database SuperCon database. Besides, from the learned features, they extract the properties of the elements and reproduce the chemical trends. In the test set performance, the RMSE and the coefficient of determination $R^2$ are 8.14 K and 0.97, respectively.\vspace{2.5pt}\\
Although studies \cite{stanev2018machine}, \cite{hamidieh2018data} and  \cite{owolabi2016application} show significant achievement in predicting the $T_c$,  each method still has limitations. For random forests, the main disadvantage is their complexity, more computational resources, less intuitive. It can quickly become inconsistent when the commonly used setups for random forests can be inconsistent. In addition, the GXBosst model is more sensitive to overfitting if the data is noisy. Consequently, with both approaches, the training generally takes longer because trees are built sequentially. Besides the advantages of SVMs, an important practical question, which is not entirely solved, is the selection of the kernel function parameters,  high algorithmic complexity, and extensive memory requirements of the required quadratic programming.\vspace{2.5pt}\\  
To overcome the remaining challenges, this paper presents a conceptual illustration of a generative Bayesian-based model. There are advantages of Bayesian model selection as follows:
\begin{enumerate}
    \item[$\bullet$] Avoid the overfitting associated  with the ML by marginalizing over the model parameters instead of making point estimates of their values.
    \item[$\bullet$] Models can be compared directly on the training data, without the need for a validation set.
    \item[$\bullet$] Avoids the multiple runs for training each model associated with cross-validation.
\end{enumerate}

Consequently, the proposed approach provides a more comprehensive and realistic picture of ML model performance in material discovery applications. Mainly, it can outperform deep learning approaches, from the study \cite{konno2018deep} and \cite{zeng2019atom}, which are claimed to be failed because of their function distributions of low cross-predictability with a descent algorithm \cite{abbe2018provable}.
\section{Model and Preliminaries}

\label{sec:model_pre}
\subsection{Bayesian Linear Regression Model}
Let us start with a simple linear regression function to approximate a true generating function such that:
\begin{equation}
    y_{i} = \mathbf{w}^{T}\mathbf{x}_{i} + \epsilon_{i}
\end{equation}

\noindent where the response $y_i$ is a linear function of the covariate $x_i \in \mathcal{R}^D$ and is linear in the parameters $w$, additional bias $\epsilon$ as well. Collecting $I$ response variable $Y, \epsilon \in \mathcal{R}^I$ we have:
\begin{equation}
    Y=XW + \epsilon
\end{equation}

\noindent where $X \in \mathcal{R}^{I \times D}, W\in \mathcal{R}^{D \times N}$ is referred to as the design matrix and weight matrix, respectively. The weight matrix is assumed the only parameter $\theta = \{W\}$. Then, the empirical cost function is as follows:
\begin{align}
\label{eq:cost_function}
    \Tilde{C}(\theta) = \frac{1}{N}\sum_{i=1}^N \frac{1}{2} \norm{y_i - W^T x_i}_2^2
\end{align}
The gradient is calculated then to find the minimum of empirical cost function from Eq. \ref{eq:cost_function}, by the following expression: 
\begin{align}
    \label{eq:grad_cost_function}
    \nabla \Tilde{C}(\theta) = -\frac{1}{N}\sum_{i=1}^N\left(y_i -W^Tx_i \right)^T x_i
\end{align}
We can use Eq. \ref{eq:cost_function} and \ref{eq:grad_cost_function} with an iterative optimization algorithm, such as gradient descent or stochastic gradient descent, to find the best $W$ that minimizes the empirical cost function, as in the study \cite{owolabi2016application}. Even though a better option is to use a validation set that can stop the optimization algorithm when the minimal validation cost function is reached. However, these methods demand clarification of a linear network because:

\begin{enumerate}
    \item[$\bullet$] First, there is no guarantee whether the real generating function $f$ is a linear function. If it is not, the linear regression model cannot be expected to approximate the true function well.
    \item[$\bullet$] Second, there is not much control over what expectations to measure the given input data $x$. Therefore, how well $x$ represents the input remains unclear. 
\end{enumerate}

Now, given a training set $\mathcal{D} = \{X,Y\}$, estimate $w$ so that the response $y^{*}$ to a new data point $x^{*}$ can be predicted by calculating the expectation $ \mathbb{E}[y^{*} | x^{*}]$ as given $ \mathbb{E}[y^{*} | x^{*}] = w^Tx^{*}$. To do that, we develop a probabilistic graphical model of the Bayesian linear regression model. Then, our target finds the likelihood function for $w$ and the prior over $w$, which is given by \cite{drugowitsch2013variational}:
\begin{equation}
\begin{aligned} p(\mathbf{y} | \mathbf{X}, \mathbf{w}) &=\prod_{i=1}^{I} \mathcal{N}\left(y_{i} | \mathbf{w}^{T} \mathbf{x}_{i}, \lambda^{-1}\right) \\  \end{aligned}
\end{equation}
\noindent where, $\lambda$ is the noise precision parameter and is assumed to be known for simplicity. Thus, the joint distribution over all the variables is given by the following factorization.
\begin{equation}
    \label{eq:learning_pro}
    p(\mathbf{y}, \mathbf{w}| \mathbf{X})=p(\mathbf{y} | \mathbf{X}, \mathbf{w}) p(\mathbf{w})
\end{equation}

The whole procedure of learning given by Eq. \ref{eq:learning_pro} is a process of searching for the best \textit{hypothesis} over the entire space $\mathcal{H}$ of hypotheses. It is assumed that each hypothesis corresponds to each possible function with a unique set of parameters and a unique functional form, and that hypothesis only takes the input $x$ and the output $y$. 

\subsection{Variational Inference for Bayesian Neural Network}
We can re-write Eq. \ref{eq:learning_pro} again with the only parameter $\theta$ for the weight matrix $W$ as the following equation. Then, the posterior inference over $w$ given by Eq. \ref{eq:eq_intract} is often intractable, especially, because of the divisor.
\begin{equation}
\label{eq:eq_intract}
p_{\theta}(w | x)=\frac{p_{\theta}(x | w) p(w)}{p_{\theta}(x)}=\frac{p_{\theta}(w, x)}{p_{\theta}(x)}=\frac{p_{\theta}(w, x)}{\int_{w} p_{\theta}(x, w)}
\end{equation}

Therefore, we assume that there is a tractable family of distribution $Q$, which is similar to $p_{\theta}(x|w)$. Then, we try to find an approximate posterior inference using $q_\phi$ ($q_\phi \in Q$). Hence, the optimization objective must measure the similarity between $p_\theta$ and $q_\phi$. To capture this, we use the Kullback-Leibler (KL) divergence as given by:
\begin{align} 
\mathrm{KL}\left(q_{\phi} \| p_{\theta}\right) &=\int_{w} q_{\phi}(w | x) \log \frac{q_{\phi}(w | x)}{p_{\theta}(w | x)} 
\end{align}

Because we cannot minimize the KL-Divergence directly, we have isolated the intractable evidence term in KL-Divergence:
\begin{align} 
\mathrm{KL}\left(q_{\phi} \| p_{\theta}\right) &=\left(\mathbb{E}_{q_{\phi}} \log \frac{q_{\phi}(w | x)}{p_{\theta}(w, x)}\right)+\log p_{\theta}(x) \nonumber \\
&=-\mathcal{L}(x ; \theta, \phi)+\log p_{\theta}(x) 
\end{align}

Then, let's us rearrange terms to express isolated intractable evidence:
\begin{equation}
\log p_{\theta}(x)=\mathrm{KL}\left(q_{\phi} \| p_{\theta}\right)+\mathcal{L}(x ; \theta, \phi) \nonumber
\end{equation}

Furthermore, KL-Divergence is non-negative, it is easily to expressed as follows:
\begin{equation}
\begin{aligned} \log p_{\theta}(x) &=\mathrm{KL}\left(q_{\phi} \| p_{\theta}\right)+\mathcal{L}(x ; \theta, \phi) \\ \log p_{\theta}(x) & \geq \mathcal{L}(x ; \theta, \phi) \end{aligned} \nonumber
\end{equation}
where
\begin{equation}
\label{eq:elbo}
\mathcal{L}(x ; \theta, \phi)=-\mathbb{E}_{q_{\phi}} \log \frac{q_{\phi}(w | x)}{p_{\theta}(w, x)}
\end{equation}

The Eq. \ref{eq:elbo} is also called the Evidence Lower Bound (ELBO). Let's us expand the derived variational lower bound, we will have then:
\begin{equation}
\resizebox{.9\hsize}{!}{
$\begin{aligned} 
\mathcal{L}(x ; \theta, \phi) &=-\mathbb{E}_{q_{\phi}}\left[\log \frac{q_{\phi}(w | x)}{p_{\theta}(w, x)}\right] \nonumber \\ 
&=\mathbb{E}_{q_{\phi}}\left[\log p_{\theta}(x | w)+\log p(w)-\log q_{\phi}(w | x)\right] \nonumber \\
\end{aligned}$}
\end{equation}
\subsection{Optimization}
The objective is to optimize the ELBO for the derived inference model, or it can be restated as the following equation:
\begin{equation}
\label{eq:final_elbo}
\mathcal{L}(x ; \theta, \phi)=\underbrace{\mathbb{E}_{q_{\phi}}\left[\log p_{\theta}(x | w)\right]}_{\text {Reconstruction likelihood }}-\underbrace{\operatorname{KL}\left(q_{\phi}(w | x) \| p(w)\right)}_{\text {divergence from prior }}
\end{equation}

Then, the gradients $\nabla_{\theta} \mathcal{L}$ and $\nabla_{\phi} \mathcal{L}$ need to be computed. To achieve that, we apply the Stochastic  Gradient Variational  Bayes  (SGVB) approach given by \cite{kingma2013auto}. Technically, the key of SGVB estimator is a reparameterization trick, i.e., they reparameterize the random variable, as given:
\begin{equation}
w \sim q_{\phi}(w | x)=\mathcal{N}\left(w | \mu_{w}(x ; \phi), \sigma_{w}^{2}(x ; \phi)\right) \nonumber
\end{equation}
as
\begin{equation}
w=w(\epsilon ; x, \phi)=\epsilon \sigma_{w}(x ; \phi)+\mu_{w}(x ; \phi), \epsilon \sim \mathcal{N}(0, I) \nonumber
\end{equation}

Then, the expectation can be written with respect to $\epsilon$:
\begin{align}
    \mathcal{L}(\phi, \theta) &=\mathbb{E}_{w \sim q_{\phi}(w | x)}\left[\log p_{\theta}(x, w)-\log q_{\phi}(w | x)\right]  \nonumber \\ 
    \begin{split}
        &=\mathbb{E}_{\epsilon \sim N(0, I)}[\log p_{\theta}(x, w(\epsilon ; x, \phi)) \\
        & \hspace{85pt} -\log q_{\phi}(w(\epsilon ; x, \phi) | x)]  \nonumber
    \end{split}
\end{align}

Consequently, the gradient with variational parameter $\phi$ can be directly moved into the expectation, enabling an unbiased low variance Monte Carlo estimator:
\begin{align} 
    \begin{split}
    \nabla_{\phi} L(\phi, \theta) &=\mathbb{E}_{\mathbf{\epsilon} \sim \mathbb{N}(0, I)} \nabla_{\phi}[\log p_{\theta}(x, w(\epsilon ; x, \phi)) \\
    & \hspace{80pt} -\log q_{\phi}(w(\epsilon ; x, \phi) | x)] \nonumber \\ 
    &\approx \frac{1}{k} \sum_{i=1}^{k} \nabla_{\phi}[\log p_{\theta}\left(x, w\left(\epsilon_{i} ; x, \phi\right)\right)\\
    & \hspace{75pt}-\log q_{\phi}\left(w\left(\epsilon_{i} ; x, \phi\right) | x\right)]
    \end{split}
\end{align}
where $\epsilon_{i} \sim \mathcal{N}(0, I)$
\section{Critical Temperature Predictive Model}
\label{sec:temp_model}
\subsection{High-Tc Data}
Although there are many public data available for superconductors \cite{wimbush2016public, supercond}, the present study used only the SuperCon database. We will restate the refined data from the study \cite{hamidieh2018data} because of significant reasons. First, the material investigated is a Standardized Data for Typical Oxide High-$T_c$ materials; all preparation, characterization is captured with i) larger amount dataset, ii) a more substantial number of features from elemental properties, iii) freely available access for everyone, and iv) compatibility as a performance benchmark. Besides, we can assess the importance of the features, which are based on thermal conductivity, atomic radius, valence, electron affinity, and atomic mass in prediction accuracy for $T_c$.

\begin{figure}[t!]
			\centering
			\includegraphics[scale=0.55]{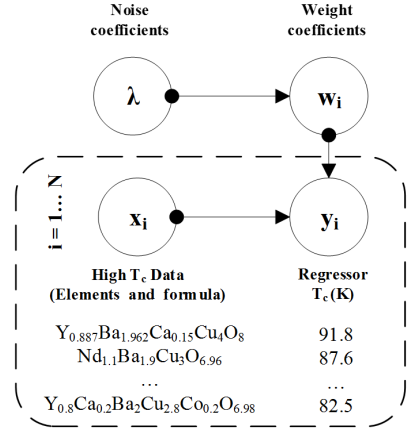}
			\caption{Probabilistic graphical model of the VBNN model to predict $T_c$.}
			\label{fig:bnnpro}
		\end{figure}

Studies \cite{stanev2018machine}, \cite{hamidieh2018data,  owolabi2016application, konno2018deep, zeng2019atom}  also create a model to predict $T_c$ from the SuperCon data. Our approach is different from those studies in the following ways: (i) We use generative neural network as illustrated in Fig. \ref{fig:bnnpro}, that probabilistic analyses and statistical learning theories are utilized to tune the learnable hyper-parameters, and (ii) most importantly, the model promises to discover rich structure (latent and distributional formation) in superconductor's chemical formula data while generating realistic data distribution from a latent code space. Then, the nature of the relationship between the features and $T_c$ can be statistically inferred from the model.
\subsection{High-Tc Prediction}
The model, in previous section \ref{sec:model_pre}, have been defined to infer the parameters. Next, the main target is to predict about new data. As consequence, the probability distribution of new data $y = T_c$ given its input feature $x$ and our training data $D$ is defined as follows:
\begin{equation}
p(y | x, D)=\int_{w} p(y | x, w) p(w | D) \nonumber
\end{equation}

Because we have learned the approximation of $p_{\theta}(w|D)$ by the variational $q_{\phi}(w)$ in Eq. \ref{eq:final_elbo}. Therefore, we can use the Monte Carlo estimation to get an unbiased estimate of it by sampling from the variational posterior as given by:
\begin{equation}
p(y | x, D) \simeq \frac{1}{M} \sum_{i=1}^{M} p\left(y | x, w\right) \nonumber
\end{equation}

As a result, the prediction for new superconductor data is the mean of the predictive distribution as expressed by: 
\begin{equation}
\hat{y}=\mathbb{E}_{p(y | x, D)} y \simeq \frac{1}{M} \sum_{i=1}^{M} \mathbb{E}_{p\left(y | x, w\right)} y 
\end{equation}
\subsection{Predictive Model Evaluation}
The most common technique for model validation is RMSE, $R^2$ and log-likelihood. RMSE is the square root of the predictive mean square error, and the smaller RMSE means, the better predictive accuracy is:
\begin{align}
    RMSE = \sqrt{\frac{1}{N}\sum_{i=1}^{N}\left( \hat{y}_i - y_i \right)^2}
\end{align}

$R^2$ values are commonly expressed as percentages from 0\% to 100\% (or its values range from 0 to 1). It approximates how well the model's input can explain the observed variation. 
\begin{align}
    R^2 = 1 - \frac{\sum_{i=1}^N(y_i - \hat{y}_i)^2}{\sum_{i=1}^N(y_i - \Bar{y}_i)^2}
\end{align}

\section{Numerical Results and Discussions}
\label{sec:result}
\begin{table}
  \caption{Numerical Result Comparison}
  \label{tb:result}
    \begin{tabular}{ccc}
    \midrule\midrule
    ML Approaches & $R^2$  & RMSE (K)            \\
     \midrule
    Random Forest \cite{stanev2018machine}  &       0.85   &   N/A             \\
    Random Forest \& XGboost \cite{hamidieh2018data}  &       0.74   & 17.6     \\ 
    Support Vector Machine \cite{owolabi2016application}   &       0.96   & N/A               \\
    Convolutional Neural Network \cite{konno2018deep}   &       0.93   & N/A               \\
    Atom Table Convolutional Neural Network  \cite{zeng2019atom}   &       \textbf{0.97}   & 8.14  \\
    Variational Bayesian Neural Network     &       0.94   & \textbf{3.83}  \\
    \midrule\midrule
    \end{tabular}
  \end{table}
For the implementation, we use the Python-based Bayesian deep learning library \cite{zhusuan2017}. Then, the predictive model training and testing are executed by using the Tensorflow on an NVIDIA Tesla k80 GPU. There is no need for the validation set to explore the effect of VBNN on overcoming the overfitting challenge, then the dataset is solely divided into 30\%, 70\% for test-set and train-set, respectively. Practically, it is problematic to determine a good network topology just from the number of inputs and outputs. Because accuracy is the main criteria for designing the VBNN, the hidden layers can be increased \cite{sheela2013review}. In general, the neural network models improve with more epochs of training, and the accuracy remains stable as they converge \cite{hoffer2017train}. And, the larger batch sizes result in faster progress in training but do not always converge as fast. In contrast, the smaller batch sizes train slower but can converge faster \cite{chen2018effect}. Besides, averaging over a multiplicity batch of 10 is going to produce a gradient that is a more reasonable approximation of the full batch-mode gradient. As a consequence, the experiment converged with the running of 1000 epochs, the batch size of 10, and 100 hidden layers. For the reproducible analysis, predictions and evaluations, the implementation code and results are available at GitHub repository (\href{https://github.com/ltdung/VBNN_HighTc}{https://github.com/ltdung/VBNN\_HighTc}). 

The presented Bayesian regression approach can also directly be applied to predict the critical temperature of a superconductor, as shown in Table \ref{tb:result}. Our confidence scores $R^2$ have strong overall concordance with previous predictions ($R^2$ = 0.94). Besides, a significant improvement was obtained in the RMSE at \textbf{3.83} K. The result is a striking illustration of VBNN performance compared with other techniques. In short, to the knowledge of the authors, the generative approach for superconductors $T_c$ prediction is the first of its kind. Our results are encouraging; however, reproducibility of replicated experiments should be conducted for worthy investigations:
\begin{enumerate}
    \item[$\bullet$] First, an important question for future studies is to use the pre-trained VBNN predictive model to validate its performance on different superconductor datasets. Possible directions are customizing the ``transfer learning'' paradigm to take advantages of the optimized hyper-parameters from the VBNN neural network.
    \item[$\bullet$] Second, future work should focus on exploring feasible compounds as a new supercondutor. It will be beneficial in having an initial feedback to determine the correctness and efficiency of alternative compounds before conducting costly, effortfully experiments in real practice.
\end{enumerate}
\section{Conclusion}
\label{sec:conclusion}
The material data science, specifically in superconductor exploring, is in the early stages of ML adoption. There is a growing number of single-use applications, but more intelligible models are yet to be seen. In this work, we developed a new probabilistic approach using variational Bayesian neural network for estimating the $T_c$ value of high-temperature superconductors.  Our results are in general agreement with existing studies in $T_c$ predictive model. These preliminary results demonstrate the feasibility of using generative neural network, which provides compelling, helpful evidence to understand the underlying superconductivity physics. This finding is promising and should be investigated with other advanced predictive models, which could eventually lead to the discovery of new superconductors in future.

\bibliographystyle{IEEEtran}

\bibliography{superconductivity}

\end{document}